# E(n)-Equivariant Cartesian Tensor Passing Potential

Junjie Wang,[1] Yong Wang,[1] Haoting Zhang,[1] Ziyang Yang,[1] Zhixin Liang,[1] Jiuyang Shi,[1] Hui-Tian Wang,[1] Dingyu Xing,[1] and Jian Sun[1,*]

[1] National Laboratory of Solid State Microstructures, School of Physics and Collaborative Innovation Center of Advanced Microstructures, Nanjing University, Nanjing 210093, China

**Abstract**

Machine learning potential (MLP) has been a popular topic in recent years for its potential to replace expensive first-principles calculations in some large systems. Meanwhile, message passing networks have gained significant attention due to their remarkable accuracy, and a wave of message passing networks based on Cartesian coordinates has emerged. However, the information of the node in these models is limited to scalars, vectors, and tensors. In this work, we proposed High-order Tensor Passing Potential (HotPP), an E(n) equivariant message passing neural network that extends the node embedding and message to an arbitrary order tensor. By performing some basic equivariant operations, high order tensors can be coupled very simply and thus the model can make direct predictions of high-order tensors such as dipole moments and polarizabilities without any modifications. Compared to high order tensor models based on spherical vectors, this network is simpler and can achieve comparable accuracy with much fewer parameters. The tests in several datasets demonstrate HotPP is a promising new approach that warrants further investigation.

---

[*] Corresponding author. jiansun@nju.edu.cn

# Introduction

Molecular dynamics (MD) is a powerful computational technique allowing for the exploration of various physical and chemical phenomena at the atomic level and the study of the behavior of molecules and materials over time. It bridges the gap between theoretical predictions and experimental observations, enabling researchers to gain a comprehensive understanding of the behavior, properties, and interactions of molecules and materials. With sufficient computational resources, first principles calculations based on Density Functional Theory (DFT)[1] can simulate systems with hundreds or even thousands of atoms. However, it struggles when it comes to larger systems. Another approach to computing atomic interactions is empirical force fields, providing much quicker calculations and the ability to handle significantly larger systems. Nevertheless, many of these force fields rely on empirical observations, limiting their applicability to specific ranges and lacking universality and transferability. The machine learning potential (MLP)[2–8], which aim to accurately describe the potential energy surface of atomic configurations, combines the advantages of both DFT and empirical force fields. A well-trained machine learning force field can achieve accuracy close to DFT and even beyond DFT[9,10], and perform very large-scale, long-time simulations, offering a glimpse into the future of research in studying complex dynamical problems.

Most existing machine learning potentials are based on the framework proposed by Behler[2], which fits the total energy as a sum of atomic energies $E = \sum E_i$, and the atomic energies are determined by the atomic environment within a certain cutoff radius. This format ensures the scalability of the potential, allowing the network to be trained on small systems and extrapolated to larger systems. The quality of such a model is highly dependent on the choice of descriptors that describe the atomic environments[11]. A reasonable descriptor should first possess invariance to rotations, translations, and atom permutation; thus, the same atomic environment yields the same atomic energy. A common approach is to construct a series of symmetric functions based on interatomic distances and angles between atoms since these two quantities are naturally invariant under rotations and translations. Depending on the number of atoms involved, a series of so-called two-body and three-body descriptors can be obtained, such as atom-centered symmetry functions (ACSF)[2,12,13], the NEP descriptor[14,15,7], the smooth overlap of atomic positions (SOAP)[16], the DeePMD descriptor[5,17]. However, these descriptors are not complete[18], as different atomic environments can yield the same descriptor. Atomic cluster expansion (ACE)[6,19] and Moment Tensor Potential (MTP)[4,20] have proposed complete descriptors that can account for interactions of arbitrary order, but the number of descriptors can easily grow to tens of thousands as the order increases. Another issue is that such descriptors are only dependent on the coordinate information within the cutoff radius. When dealing with long-range interactions, simply increasing the cutoff radius would significantly raise computational complexity since the number of atoms is proportional to the cube of cutoff radius.

Message passing network (MPN)[21] can help address both of these issues. In the context of MLP, MPN is used to represent molecule or crystal structures as graphs, where atoms are nodes and bonds are edges. The key idea behind MPN is the iterative passing of messages between nodes, allowing information to be exchanged and aggregated. Such message passing processes can, on one hand, lead to the emergence of multiple atoms in the final descriptor (thus, resulting in n-body symmetric functions). On the other hand, it allows information from atoms beyond the cutoff radius to be transmitted to the current atom. As a result, many machine learning potentials based on message passing networks have achieved high levels of accuracy[22–28,8,29,30]. It is worth noting that as long as the energy obtained at the end satisfies the symmetry requirements, the messages do not necessarily have to be scalars. For example, NequIP[8], BotNet[29], and MACE[30] utilize high-order tensors based on spherical harmonics in the message passing, coupling them through Clebsch-Gordon (CG) coefficients to construct equivariant networks. These methods have shown significant improvements in accuracy compared to approaches that only use scalar messages. Another category of methods, including PaiNN[25] and torchMD-Net[27], directly utilize vectors in Cartesian space as messages and obtain equivariant results through a series of designed layers. This approach does not require coupling through CG coefficients, but Cartesian vectors is only equivalent to $l=1$ tensors in the spherical harmonics method. TeaNet[31] can pass matrices information equivalent to $l=2$ tensors, but this introduces a multitude of artificially designed duplications, resulting in a highly intricate network structure that becomes challenging to extend to higher order tensors. Networks that use tensors of arbitrary orders as messages based on Cartesian coordinates have not been proposed.

In this work, we proposed High-order Tensor Passing Potential (HotPP), which can utilize arbitrary order Cartesian tensors as messages. By combining some basic equivariant operations between tensors, all the high-order tensors used in the network are E(n)-equivariant, thus the output is consistent with the rotation of coordinates. In other words, if the output is a scalar, it remains invariant under rotations, while the vector output will rotate in accordance with the rotation of the coordinates, and the matrix output will transform as $M' = RMR^T$. Therefore, the method can directly predict high-order tensors such as dipole moments and polarizability tensors without any modifications. We valid HotPP in three prediction tasks: the energies and forces of molecular dynamics trajectory of small molecule; the energies, forces, and stresses of carbon with periodic boundary conditions; and the dipole moments and polarizability tensors of small molecules with coupled cluster singles and doubles (CCSD) accuracy. In these tests, our model achieves good performance with fewer parameters comparable to other high-order models, which provide a novel frame of equivariant network based on Cartesian coordinates.

**Equivariant functions of Cartesian tensors**

Cartesian tensors are the tensors that transform under rotations in Euclidean space in a simple way. In other words, a Cartesian tensor is a tensor whose components transform as a product of vectors and covectors under rotations, without any additional factors that depend on the rotation matrix itself. Specifically, a n-th rank tensor transforms as:

$$T_{i_1 i_2 \cdots i_n} \to T'_{i_1 i_2 \cdots i_n} = (R_{i_1 j_1})(R_{i_2 j_2}) \cdots (R_{i_n j_n}) T_{j_1 j_2 \cdots j_n} \tag{1}$$

where R is an orthogonal matrix. Under this definition, it is easy to find that since $v_i \to v'_i = R_{ij} v_j$, the vectors are first-order tensors. And the dyadic product of two vectors is a second-order tensor because $(uv)_{i_1 i_2} \to (uv)'_{i_1 i_2} = (R_{i_1 j_1})(R_{i_2 j_2})(uv)_{j_1 j_2}$.

And equivariance is a property of functions or transformations between two spaces, where the transformation preserves the relationships between the elements of those spaces. More formally, a function $\phi: X \to Y$ is said to be equivariant with respect to a group G acting on two sets X and Y if for all $g \in G$ and $x \in X$, we have:

$$\phi(g \circ x) = g \circ \phi(x) \qquad (2)$$

This means that applying a function $\phi$ to an object x and then applying a group element $g$ to the resulting object should give the same result as first applying the group element to the object and then applying the function. And the composition of equivariant maps is also equivariant:

$$\psi(\phi(g \circ x)) = \psi(g \circ \phi(x)) = g \circ \psi(\phi(x)) \qquad (3)$$

Therefore, by providing some basic equivariant functions between Cartesian tensors and combining them, we can obtain an equivariant neural network. Here, we use the following three equivariant operations:

1. Linear combinations of tensors with the same order: $f(T_1, T_2, \cdots, T_m) = \sum c_i T_i$.

2. Contraction of two tensors.

    The contraction of tensors is a mathematical operation that reduces the rank of tensors by summing over one or more pairs of indices. For example, consider a 3-order tensor A and a 2-order tensor B, the contraction of them can be $C_{ijk} = A_{ijl} B_{kl}$, this will reduce the rank of tensors by 2. More generally, if we sum over more than one pair of indices between an x-order tensor $T_1$ and a y-order tensor $T_2$ such as:

$$\phi_z(T_1, T_2)_{a_1 \cdots a_{x-z} b_1 \cdots b_{y-z}} = T^1_{a_1 \cdots a_{x-z} c_1 \cdots c_z} \cdot T^2_{b_1 \cdots b_{y-z} c_1 \cdots c_z} \qquad (4)$$

    We can get a new tensor with $x + y - 2z$ order, where $0 \leq z \leq \max(x, y)$. When z=0, none of the indices and the equation (4) becomes tensor product $T^{1 \otimes 2}_{a_1 \cdots a_x b_1 \cdots b_y} = T^1_{a_1 \cdots a_x} \cdot T^2_{b_1 \cdots b_y}$.

3. Partial derivative with respect to another Cartesian tensor: $\frac{\partial}{\partial T_{j_1 j_2 \cdots j_n}}$.

By the combination of these operations, we can get many equivariant functions. Many common operations frequently used in equivariant neural networks that operate on vectors, such as scaling of vectors: $s \cdot \vec{v}$, scalar products $\langle \vec{v_1}, \vec{v_2} \rangle$, vector products $\vec{v_1} \times \vec{v_2}$ can all be viewed as special cases of these three operations. Some other more complex descriptors, such as MTP descriptors[4,20], can also be obtained through combinations of these operations as shown in the Supplementary Information.

**Equivariant message passing neural network**

To obtain an end-to-end machine learning model for predicting material properties, the input should be the positions $\{r_i\}$ and chemical elements $\{Z_i\}$ of all atoms, and for periodic crystals, the lattice parameters should also be considered. To apply graph neural networks, we first transform the crystal structure into a graph $\{n_i, e_{ij}\}$, where each node $n_i$ corresponds to an atom $i$ in the unit cell, and all atoms $j$ within a given cutoff distance $r_{cut}$ are considered connected to $n_i$ labeled with their relative positions $r_{ij}$. For periodic structures, since atom $j$ and its equivalent atom $j'$ may both lie within the cutoff distance of atom $i$, $n_i$ may have more than one edge connected to $n_j$. To extract the information of the nodes, we use the scheme of MPN. A normal MPN can be described as:

$$m_i^{t+1} = \bigoplus_{j \in N(i)} M_t(h_i^t, h_j^t, e_{ij}) \tag{5}$$

$$h_i^{t+1} = U_t(h_i^t, m_i^{t+1}) \tag{6}$$

where $h_i^t$ is the hidden feature of $n_i$ at layer t that captures its local information, messages are then passed between nodes along edges, with the message at each edge $e_{ij}$ being a function $M_t$ of the features of the nodes connected by that edge. The $\oplus$ is a differentiable and permutation invariant function such as sum, mean or max to aggregate the message at each node together to produce an updated message $m_i^{t+1}$ for that node, which in turn is used to update the hidden feature with the function $U_t$ for the next iteration.

A concrete example illustrating the principles of MPN is presented in Fig 1. We first determine the key connectivity of a structure based on a given cutoff radius and convert it into a graph. Then the messages on the nodes can be passed through the edges by a two-body interaction. As the process of message passing, the information from atoms beyond the cutoff radius can also be conveyed to the central atom. As illustrated in Fig.1(b), the blue arrows represent the first time of message passing, while the green arrows denote the second. During the initial time of message passing, information of atom 1 is encoded into the hidden information of atom 2. Subsequently, in the second time of message passing, the information of atom 2 including some information of atom 1 is collectively transmitted to atom 3, thereby achieving non-local effects from atom 1 to atom 3. On the other hand, due to the interaction between atom 4 and atom 2 in the second time of message passing also containing the information from atom 1, the effective interaction is elevated from a two-body interaction to a three-body interaction. This indeed encapsulates the two advantages of the message-passing architecture.

However, the scalar hidden feature, message, and the edge information (always relatively distance) here will limit the expressive capacity and may cause the incompleteness of atomic structure representations. As shown in Fig.2(a) and Fig.2(d), if we only use scalar information $h_i$, $h_j$, and $d_{ij}$ to pass the message in equation (5) and update the feature in equation (6), all nodes will always produce the same embedding information. As a result, the network will be unable to distinguish between these two structures and give the same total energy. Even if the 3-body message includes angles $a_{ijk}$ are taken into consideration, some structures with only 4 atoms cannot be distinguished[25], as shown in Fig.2(b) and Fig.2(e). Due to the identical atomic environments within the truncation radius, no matter how many message passing iterations are performed, these two different structures will only yield the same result. To alleviate this problem, a series of models that use high-order geometric tensors during the message passing have been proposed. For example, allowing vectors in the message passing process can differentiate Fig.2(b) and Fig.2(e), but in the case of Fig.2(c) and Fig.2(f), the summation in equation (5) would cause the network to confuse these two structures. It can be anticipated that increasing the order of tensors in message passing would enhance the expressive power of the network. Previously, the order of high-order tensor networks based on Cartesian space was typically limited to 2, while our method can work with any order Cartesian tensors. In the following, we use $^l h_i^t$ to represent the $l$-order Cartesian tensor features of node $i$ in the t-th layer, and $r^{\otimes n}$ to represent tensor product of a vector $r$ for n times: $r \otimes r \otimes \cdots \otimes r$. In particular, for n=0 we define this to a learnable function of $\|r\|$.

**Initialize of node features.** The scalar features in the first layer $^0h_i^0$ should be invariant to rotation, translation, and permutation of the atoms with the same chemical species. This is also the requirement for most descriptors used in machine learning potentials, so these descriptors such as ACSF, SOAP, ACE, MTP, etc., can be used directly to expedite the process of feature extraction. Here, we used the trainable chemical embedding similar to SchNet[22] to minimize human-designed elements as much as possible., that is, for each element $Z_i$ there is a learnable vector. For high-order features $^l h_i^0$ with $l > 0$, we set them all to 0 at the beginning, and $\sum_{j \in N(i)} r_{ij}^{\otimes l}$ can be another choice.

**Message and aggregate.** In order to combine the information of neighboring nodes, we need to design a message passing function $M_t$ in equation (5). Considering that the hidden feature $^{l_i}h_i^t$, $^{l_j}h_j^t$, the bond info $e_{ij}$, and the target message $^{l_{out}}m_{ij}^{t+1}$ can be tensors of arbitrary order. Therefore, we need to find an equivariant way to compose the two tensors to a new tensor with different order, and equation (4) is such an operation. In our model, we write $M_t$ in equation (5) as:

$$^{l_o}_{l_i,l_r}m_{ij}^t = M_t(h_i^t, h_j^t, e_{ij}) = f_{l_r}^t(d_{ij}) \cdot {}^{l_i}h_j^t{}_{a_1 \cdots a_{l_i-l_c}c_1 \cdots c_{l_c}} \cdot \left(u_{ij}^{\otimes l_r}\right)_{c_1 \cdots c_{l_c} b_1 \cdots b_{l_r-l_c}} \quad (7)$$

Where $d_{ij} = \|r_{ij}\|$ is the relative distance between atom i and atom j, $u_{ij} = \frac{r_{ij}}{d_{ij}}$ is the direction vector, $0 \le l_c \le \min(l_i, l_r)$ is the number of the indices summing up during the contraction. $f_{l_r}^t(d_{ij})$ is the radial function, which is a learnable multi-layer perceptron of radial basis functions such as Bessel basis or Chebyshev basis. The result is a Cartesian tensor with order $l_o = |l_i + l_r - 2l_c|$, which is between $|l_i - l_r|$ and $l_i + l_r$. Since $l_r$ can be chosen arbitrarily, we can obtain an equivariant tensor of order from 0 to any arbitrary order.

We use a summation operation as the aggregation function for the messages in equation (5), that is, directly adding all the messages obtained from neighboring nodes. For tensors of the same order obtained from different $(l_i, l_r)$, we add them together with different coefficients. Due to the arbitrariness of $l_o$ and $l_r$, we need to specify their maximum values. With given $l^t_{omax}$, $l^t_{rmax}$, we sum up all possible $(l_i, l_r)$:

$$^{l_o}m_i^{t+1} = \sum_{l_r \leq l^t_{rmax}} \sum_{l_i} c_{l_o,l_r,l_i} \sum_{j \in N(i)} f_{l_r}(d_{ij}) \cdot {}^{l_i}h^t_{j\,a_1\cdots a_{l_i-l_c}c_1\cdots c_{l_c}} \cdot \left(u^{\otimes l_r}_{ij}\right)_{c_1\cdots c_{l_c}b_1\cdots b_{l_r-l_c}} \quad (8)$$

**Update.** For scalar message $^0m_i^{t+1}$, we feed it to a fully connected layer followed by a non-linear activation function to extract the information, and update the hidden feature with the structure similar to residual neural networks:

$$^0h_i^{t+1} = {}^0h_i^t + \sigma({}^0W^t \cdot {}^0m_i^{t+1} + {}^0b^t) \quad (9)$$

Where $\sigma$ is the nonlinear activation function, $^0W^t$ and $^0b^t$ are the weights and bias in the t layer for the scalar message. However, for the tensors above 0 order, both the bias and the activation function will break the equivariance. Therefore, we only apply bias when $l = 0$.

For the high order activation function, as shown in equation (3), tensor multiplication by a scalar is equivariant. Hence, we need to find a mapping from an n-order tensor to a scalar. One simple idea is to use the squared norm of the tensor $\|x\|^2 = \sum_{a_1,\cdots,a_l} x^2_{a_1\cdots a_l}$ since it is equivariant to E(n) by definition. Therefore, for $l > 0$, we write the element-wise non-linear function as:

$$\sigma^l(x_{ia_1\cdots a_l}) = \sigma'(\|x\|^2 + b_l) \cdot x_{ia_1\cdots a_l} \quad (10)$$

It should be noted that different notations were used for the activation function in equations (9) and (10), as the choice of activation function may vary for scalar and high-order tensors. Take the SiLU function $\sigma(x) = x \cdot sigmoid(x)$ for example. For scalar, SiLU maps $x$ to $x$ itself when $x \gg 0$. However, for higher-order tensors, equation (13) will map $x_{ia_1\cdots a_l}$ to $X \cdot x_{ia_1\cdots a_l}$ instead of $x_{ia_1\cdots a_l}$ when $x_{ia_1\cdots a_l} \gg 0$. This is because if we apply the formula for higher-order tensors to scalar, which is $y_i = \sigma(x_i + b) \cdot x_i$, an extra $x_i$ is multiplied. Therefore, if we use SiLU function for $\sigma$, we should use Sigmoid function for $\sigma'$. Other activation functions can be handled using a similar approach. And hence the update function for high-order tensors is:

$$^lh_i^{t+1} = {}^lh_i^t + \sigma^l({}^lW^t \cdot {}^lm_i^{t+1}) \quad (11)$$

**Readout.** For a target n-order property, we utilize a two-layer nonlinear MLP to operate on the n-order tensor at the last hidden layer. For the same reason, bias and element-wise nonlinear functions cannot be used in the high order tensor.

$$^l o_i = \begin{cases} ^0W_2 \cdot \sigma(^0W_1 \cdot {}^0h_i^t + {}^0b^t) + {}^0b^t, & l = 0 \\ ^lW_2 \cdot \sigma^l(^lW_1 \cdot {}^lh_i^t), & l > 0 \end{cases} \quad (12)$$

**Results**

We validate the accuracy of our method on a diverse range of systems, including small organic molecules, periodic structures, and predictions of dipole moments and polarizability tensor. For each system, we trained HotPP model on two different datasets and compared the results with other models. To demonstrate the robustness of our model, all datasets were trained using the same network architecture and almost the same hyperparameters as shown in Fig.3. We use 4 message passing layers to propagate the information between atoms, the max order of the tensor of the out layer $l_{omax}^t$ and the tensor product of the relative coordinates $l_{rmax}^t$ are both set to 2 (matrix). The radial Bessel function $B(d_{ij}) = \frac{2}{r_c} \cdot \frac{\sin\left(\frac{b\pi}{r_c}d_{ij}\right)}{d_{ij}}$ are used as the basis function and the radial function $f_{l,r}^t(d_{ij})$ in Equation (8) is obtained by a 3 layer perceptron with 64 nodes in hidden layers, the Bessel roots $b$ are initialized with 1 to 8 and are optimized during the training following NequIP[8].

**Small organic molecule.** We first test our model on molecular dynamics trajectories of small organic molecules. The ANI-1x dataset[34,35] contains DFT calculations for approximately five million diverse molecular conformations obtained through an active learning algorithm. To evaluate the extrapolation capability of HotPP, we train our model with 70% data from the ANI-1x dataset and test on the COmprehensive Machine-learning Potential (COMP6) benchmark[34], which samples the chemical space of molecules larger than those included in the training set. The results are shown in Table I. Compared to ani-1x, our model has demonstrated superior performance across the majority of prediction tasks.

**Periodic systems.** After testing HotPP on small molecule datasets without periodicity, we evaluated its performance on periodic systems. We selected the carbon system with various phases as the first example[36]. It is a complicated dataset with a wide range of structures containing structural snapshots from ab initio MD and iteratively extended from GAP-driven simulations, and randomly distorted unit cells of the crystalline allotropes, diamond, and graphite. We show the results in Table II. Clearly, our model demonstrates a significant advantage in predicting forces and virial compared to most models, and only trailing slightly behind the NequIP model with the tensor rank $l$=3. However, on one hand, since the NequIP cannot predict virial, the emphasis on energy in the loss function may be greater. On the other hand, the $l$=3 NequIP model has around 2M parameters and ours utilizes only about 160k parameters. So we consider such a discrepancy to be acceptable. Overall, HotPP performs significantly well on this dataset.

Next, we verify the accuracy of our potential in calculating phonon dispersions of diamond, which was not well-predicted in some of the previous models for carbon[37]. We can obtain the force constant matrix of the structure directly through automatic differentiation $\frac{\partial^2 E}{\partial r_{i\alpha} \partial r_{j\beta}}$. We used Phonopy Python package[38,39] to calculate the phonon spectrum of diamond and compared it with the results from DFT in Fig 4. The results show that HotPP can describe the vibrational behavior well. Although there are relatively large errors in the high-frequency part at the gamma point, this could be attributed to the inaccuracy of the DFT calculations within the training dataset. We retrained the model using a more accurate dataset[37], and the newly calculated phonon spectrum almost perfectly matches the results from DFT, which demonstrates the reliability of our model.

**Dipole moment and polarizability tensor**

Since our model can directly output vectors, we attempted to directly predict the dipole moments and polarizability tensor of structures in the final section. We consider the water systems including water monomer, water dimer, Zundel cation, and liquid water bulk[41]. The dipole and polarizability of the aperiodic systems were calculated by CCSD theory and those of liquid water were calculated by DFT. Each system contains 1000 structures and we use 70% of them as training set and the rest as testing set. We calculate the RMSEs relative to the standard deviation of the testing samples to compare with previous results obtained by other models[41–43] as shown in Table III. In most cases, HotPP gets the best results except for the polarizability tensor of the water monomer. Compared to T-EANN[42] and REANN[43], HotPP performs particularly well in the case of the dipole moment of liquid water. This may be because they fit the dipole moment by learning $q_i$ and calculating $\vec{\mu} = \sum_{i=1}^{N} q_i \vec{r_i}$, which is inappropriate for periodic systems. In contrast, the output results of our model are all obtained through relative coordinates and thus we can get rid of the selecting of reference point.

Since now we can obtain the dipole moment and polarizability by HotPP, we can calculate the infrared (IR) absorption spectrum and Raman spectrum for liquid water. We separately trained a machine learning potential to learn the energy, forces, and stresses of liquid water to assist us in conducting dynamic simulations. With this potential, we perform a classical MD simulation under ambient conditions (300K, 1 bar) for 100ps and calculate the dipole moment and the polarizability tensor every 1fs. Then we compute the IR and Raman spectra by Fourier transforming the autocorrelation function (ACF) of them and the results are compared to the experiment data[44,45] as shown in Fig.5. We can observe that both HotPP model and DeePMD model can closely approximate the experimental IR spectra. Our results accurately fit the first three peaks, corresponding to the hindered translation, libration, and H-O-H bending respectively, but there is a long tail in comparison to the experimental data for the O-H stretching mode. This discrepancy may arise from not accounting for quantum effects in our classical molecule dynamic simulation. And for Raman spectra, our model also gives the result in very good agreement with experimental data.

## Conclusion

In this work, we introduce HotPP, a novel E(n) equivariant high-order tensor message passing network based directly on Cartesian space. Compared to other Cartesian-space based high-order tensor networks, HotPP can utilize tensors of arbitrary order, providing enhanced expressive power. In contrast to high-order tensor networks based on spherical harmonics and coupled with CG coefficients, HotPP employs simple tensor contraction operations, resulting in a significantly reduced number of parameters (by one or two orders of magnitude). Moreover, the network's output can be any-order tensor, enabling convenient prediction of vector or tensor properties. With its ability to achieve high accuracy while saving substantial computational time compared to first-principles calculations, HotPP holds great promise in scenarios such as molecular dynamics simulations and structure optimizations, where exploration of potential energy landscapes is essential. In future work, we would investigate approaches to eliminate redundancies in high-order Cartesian tensors to further enhance accuracy. Additionally, due to its E(n) equivariance (rather than E(3)), HotPP can be explored for high-dimensional structure optimization[46] to expedite potential energy surface exploration. It can also serve as a foundation for generating models to directly generate structures or predict wave functions. Overall, HotPP is a promising neural network that we believe can facilitate further explorations in physical chemistry, biology, and related fields.

# Figures

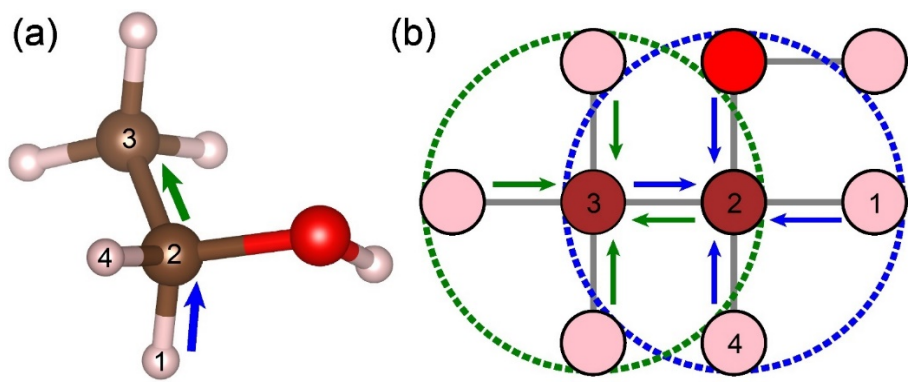

**Figure 1 The schematic diagram of message-passing networks.** (a) the origin structure; (b) the graph corresponding (a). The blue arrows represent the first time of message passing, while the green arrows denote the second.

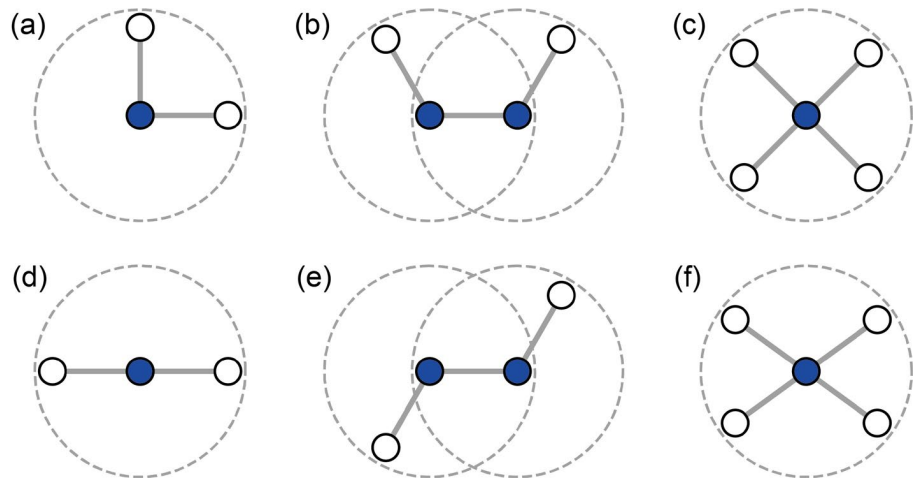

**Figure 2 Some structures that cannot be distinguished by message passing networks that do not utilize high-order tensor information.** (a) (d) cannot be distinguished by using two-body scalar information; (b) (e) cannot be distinguished by using three-body scalar information; and (c) (f) cannot be distinguished by using two-body vector information.

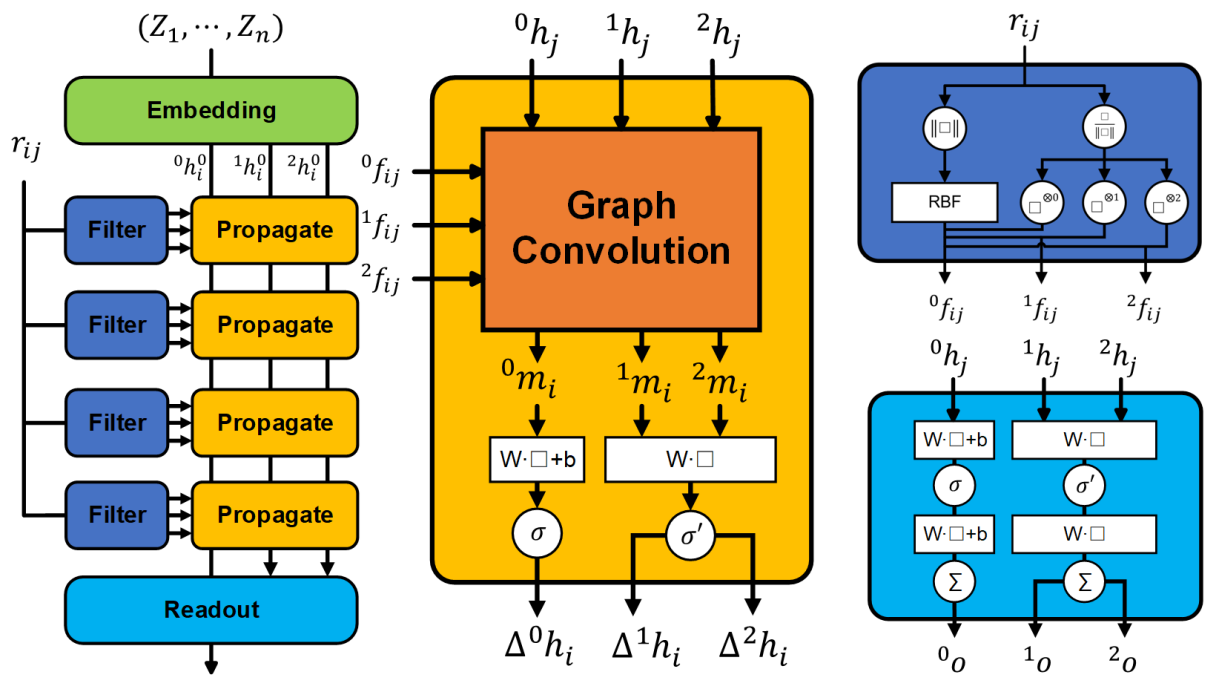

**Figure 3 The architecture of HotPP.** After embedding atomic information into scalars, vectors, and tensors, 4 propagation layers are used to further extract information from it. The final output is generated through the readout layer. For higher-order tensors or additional propagation layers, the frame is similar. Transparent boxes represent the inputs of the function. $\sigma'$ is the activation function of tensors described in equation (10). $r_{ij}$ is the distance vector between two atoms, $^n f_{ij}$ is the filter tensor in equation (8), and the $^n o$ is the final output.

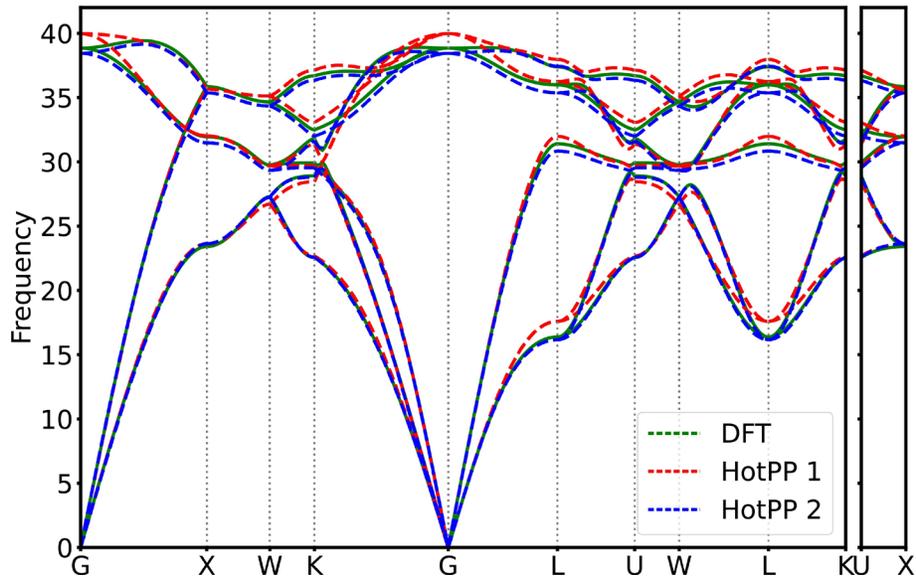

**Figure 4 Phonon spectrum of diamond.** Phonon dispersion relation for diamond as predicted by HotPP trained with the dataset 2017[36] (blue dotted) and 2020[37] (red dotted) with comparison to DFT reference data (green solid).

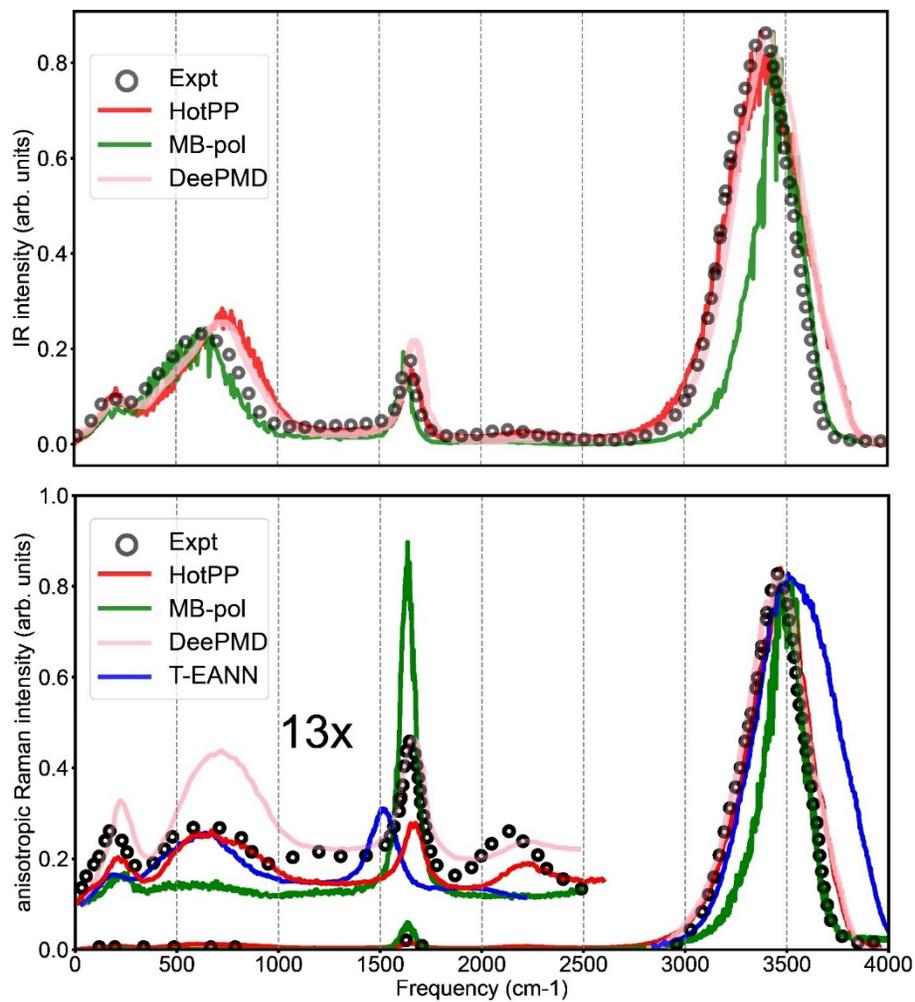

**Figure 5** Experimental and simulated IR and reduced anisotorpic Raman spectra of liquid water under ambient condition.

**Table I test results on COMP6 dataset.** The upper and lower portions of each row in the table correspond to energy and forces mean absolute errors in meV and meV/Å. The models are trained on part of the Ani-1x dataset and tested on the COMP6 benchmarks to evaluate the extrapolation capability of HotPP and Ani-1x.

|  |  | *ANI-MD* | *DrugBank* | *GDB7-9* | *GDB10-13* | *s66x8* | *Tripeptides* |
|---|---|---|---|---|---|---|---|
| *HotPP* | E | 1155 | 534 | 209 | 332 | 231 | 311 |
|  | F | 61 | 62 | 61 | 92 | 32 | 51 |
| *Ani-1x*[34] | E | 2249 | 597 | 56 | 134 | 568 | 563 |
|  | F | 249 | 174 | 108 | 170 | 136 | 149 |

**Table II test results on Carbon.** E, F, and W are RMSE of energy, forces, and virial in meV/atom, meV/Å, and meV/atom respectively. For MTP, REANN, and NEP, we only show the models that can deliver the best results (se2+se3 for DP, 5 Å cutoff for MTP, 4 Å cutoff for MTP, and 4.2 Å cutoff for NEP). More details of training parameters can be found in Fan's previous work[7].

|   | GAP[7] | DP[7] | MTP[7] | REANN[7] | NEP[7] | NequIP (l=1) | NequIP (l=2) | NequIP (l=3) | HotPP |
|---|---|---|---|---|---|---|---|---|---|
| E | 46 | 44 | 35 | 31 | 42 | 67 | 23 | 17 | 22 |
| F | 1100 | 800 | 630 | 640 | 690 | 746 | 507 | 431 | 439 |
| W | - | 170 | 200 | - | 160 | - | - | - | 64 |

**Table III Comparison the results obtained by SA-GPR, T-EANN, REANN and HotPP in different water systems.** In each system, the first line represents the relative RMSE of the dipole moment, while the second line indicates the relative RMSE of the polarizability tensor. For SA-GPR, isotropic and anisotropic terms are learned separately, and the values of liquid water are used for the qualitative comparison only as mentioned in the work of REANN[43].

|  | SA-GPR[41] | T-EANN[42] | REANN[43] | HotPP |
|---|---|---|---|---|
| $H_2O$ | ~0.11 | 0.02 | 0.05 | 0.02 |
|  | ~0.02/0.12 | 0.02 | 0.06 | 0.05 |
| $(H_2O)_2$ | ~5.3 | 6.6 | 3.0 | 2.36 |
|  | ~6.4/7.8 | 4.2 | 1.6 | 0.99 |
| $H_5O_2^+$ | ~2.4 | 1.3 | 0.4 | 0.15 |
|  | ~3.8/0.97 | 0.3 | 0.1 | 0.09 |
| Liquid water | - | 16 | 15 | 0.70 |
|  | ~5.8/19 | 2.2 | 2.1 | 0.48 |